\documentclass[sn-apa]{sn-jnl}

\usepackage{dirtytalk}
\usepackage{fontawesome5}
\usepackage{graphicx}%
\usepackage{multirow}%
\usepackage{amsmath,amssymb,amsfonts}%
\usepackage{amsthm}%
\usepackage{mathrsfs}%
\usepackage[title]{appendix}%
\usepackage{xcolor}%
\usepackage{textcomp}%
\usepackage{manyfoot}%
\usepackage{booktabs}%
\usepackage{algorithm}%
\usepackage{algorithmicx}%
\usepackage{algpseudocode}%
\usepackage{listings}%
\usepackage{soul}
\usepackage{array}
\usepackage{rotating} 

\theoremstyle{thmstyleone}%
\theoremstyle{thmstyletwo}%

\theoremstyle{thmstylethree}%

\begin{document}

\title{Lessons Learned from Designing an Open-Source Automated Feedback System for STEM Education}

\author*[1]{\fnm{Steffen} \sur{Steinert\orcidcustom{0000-0001-6364-934X}} }\email{steinert.steffen@physik.uni-muenchen.de}
\equalcont{These authors contributed equally to this work.}

\author*[2,3]{\fnm{Lars} \sur{Krupp\orcidcustom{0000-0001-6294-2915}}}
\email{lars.krupp@dfki.de}
\equalcont{These authors contributed equally to this work.}

\author*[4]{\fnm{Karina E.} \sur{Avila\orcidcustom{0000-0002-2087-3592}}}
\email{kavila@rhrk.uni-kl.de}
\equalcont{These authors contributed equally to this work.}

\author[5]{\fnm{Anke S.} \sur{Janssen\orcidcustom{0000-0003-1701-1876}}}

\author[1]{\fnm{Verena} \sur{Ruf}}

\author[1]{\fnm{David} \sur{Dzsotjan}}

\author[6]{\fnm{Christian} \sur{De Schryver}}

\author[2,3]{\fnm{Jakob} \sur{Karolus}}

\author[4]{\fnm{Stefan} \sur{Ruzika}}

\author[5]{\fnm{Karen} \sur{Joisten}}

\author[2,3]{\fnm{Paul} \sur{Lukowicz}}

\author[1]{\fnm{Jochen} \sur{Kuhn}}

\author[6]{\fnm{Norbert} \sur{Wehn}}

\author[1]{\fnm{Stefan} \sur{Küchemann\orcidcustom{0000-0003-2729-1592}}}

\affil*[1]{\orgdiv{Faculty of Physics, Chair of Physics Education}, \orgname{Ludwig-Maximilians-University Munich}, \orgaddress{ \city{Munich}, \postcode{80539}, \country{Germany}}}

\affil[2]{\orgdiv{Embedded Intelligence}, \orgname{German Research Center for Artificial Intelligence}, \orgaddress{ \city{Kaiserslautern}, \postcode{67663}, \country{Germany}}}

\affil[3]{\orgdiv{Department of Computer Science}, \orgname{RPTU Kaiserslautern-Landau}, \orgaddress{\city{Kaiserslautern}, \postcode{67663} \country{Germany}}}

\affil[4]{\orgdiv{Department of Mathematics}, \orgname{RPTU Kaiserslautern-Landau}, \orgaddress{\city{Kaiserslautern}, \postcode{67663}, \country{Germany}}}

\affil[5]{\orgdiv{Department of Philosophy}, \orgname{RPTU Kaiserslautern-Landau}, \orgaddress{\city{Kaiserslautern}, \postcode{67663}, \country{Germany}}}

\affil[6]{\orgdiv{Department of Electrical and Computer Engineering}, \orgname{RPTU Kaiserslautern-Landau}, \orgaddress{ \city{Kaiserslautern}, \postcode{67663},\country{Germany}}}



\definecolor{idcolor}{HTML}{A6CE39}
\newcommand{\orcidcustom}[1]{\href{https://orcid.org/#1}{\color{idcolor} \textsuperscript{\faOrcid}}}


\abstract{As distance learning becomes increasingly important and artificial intelligence tools continue to advance, automated systems for individual learning have attracted significant attention. However, the scarcity of open-source online tools that are capable of providing personalized feedback has restricted the widespread implementation of research-based feedback systems.
In this work, we present RATsApp, an open-source automated feedback system (AFS) that incorporates research-based features such as formative feedback. The system focuses on core STEM competencies such as mathematical competence, representational competence, and data literacy. It also allows lecturers to monitor students' progress. We conducted a survey based on the technology acceptance model (TAM2) among a set of students (N=64). Our findings confirm the applicability of the TAM2 framework, revealing that factors such as the relevance of the studies, output quality, and ease of use significantly influence the perceived usefulness. We also found a linear relation between the perceived usefulness and the intention to use, which in turn is a significant predictor of the frequency of use. 
Moreover, the formative feedback feature of RATsApp received positive feedback, indicating its potential as an educational tool. 
Furthermore, as an open-source platform, RATsApp encourages public contributions to its ongoing development, fostering a collaborative approach to improve educational tools.}




\maketitle

\section{Introduction}\label{introduction}

The demand for online learning opportunities is rising due to enhanced accessibility and interconnection~\citep{harasim2000shift,anderson2003modes}. Even before the onset of the COVID-19 pandemic, which triggered an increase in demand for online learning materials, a rising number of students turned to online resources to supplement their learning~\citep{chandra2021online,muthuprasad2021students}. 
To address this demand, a variety of applications are developed to support students in their educational journey. One such type of system is an Automated Feedback System (AFS), which offers students immediate and customized instructions or feedback. AFS encompasses a broad range of educational tools, including Intelligent Tutoring Systems (ITS), teaching platforms, or online feedback tools~\citep{deeva2021review,steinert2023harnessing}.
These systems share a common goal: to provide feedback that facilitates students' learning and improvement. One type of feedback of AFSs is formative feedback~\citep{kulhavy1989feedback, butler1995feedback,hattie2007power}, which is intended to modify the students' thinking or behavior to enhance their learning~\citep{shute2008focus}. 

In a world that is increasingly determined by data, mathematical skills are considered one of the key competencies of the 21st century~\citep{pisa2022mathematics}. These skills are particularly important for students pursuing studies in science, technology, engineering, and mathematics (STEM)~\citep{blumenthal1961multiple, hudson1982combined, hake2002relationship}. However, many students struggle with developing mathematical proficiency and these difficulties can have negative impact on their performance in other subjects \citep{ballard2004basic,pozo2006requiring,llamas2012mathematical}. This highlights the need for effective educational interventions aimed at facilitating the development of students’ mathematical skills and enhancing teachers’ pedagogical methods. Such interventions may include technological-enhanced learning tools such as e-learning platforms. 

There are numerous AFSs available for STEM education. Each system has its own level of quality, functionality, and support. While it is beyond the scope of this article to discuss all of them, interested readers can refer to existing reviews on the subject \citep{deeva2021review, mousavinasab2021intelligent}. In their review of AFSs, \cite{deeva2021review} found a shortage of open-source and well-maintained systems. During the writing of this manuscript, a new open-source web-based adaptive tutoring system called OATutor was published~\citep{pardos2023oatutor}. It includes knowledge tracing, an A/B testing framework, and LTI support. Another example of an open-source adaptive system is COLT~\citep{myers2017implementing}, which is web-based and designed to improve coding literacy. However, it has fewer features compared to OATutor.

In this article, we introduce RATsApp (Rapid Assessment Tasks App), an AFS developed by an interdisciplinary team of computer scientists, engineers, ethicists, as well as mathematics and physics education researchers. RATsApp is an open-source web-based application that combines the theoretical framework of formative feedback~\citep{kulhavy1989feedback, butler1995feedback,hattie2007power} with learning scaffolds and an evaluation of three key STEM competencies: mathematical competence~\citep{pisa2022mathematics}, representational competence~\citep{kuchemann2021inventory}, and data literacy~\citep{schuller2019future,kippers2018data}. Grounded in the aforementioned educational research, the system is designed to facilitate STEM learning by providing students with the opportunity to solve rapid assessment tasks (RATs). RATsApp employs both data-driven and expert-driven feedback to improve students' conceptual understanding.
In addition to enhancing conceptual understanding, the system also aims to help students improve the aforementioned key STEM competencies. Additionally, the system provides a dashboard to students that displays their overall performance and their progress in the STEM competence levels. A lecturer dashboard is also available, allowing lecturers to monitor their lectures and track the progress of their students. Furthermore, RATsApp was developed in accordance with existing ethical guidelines for the development of digital technologies and systems~\citep{hleg2019ethics, Montreal2018, IEEE2021}.

RATsApp is designed to be easily expandable with machine-learning algorithms for automatic question selection, adaptive feedback, and provision of helpful materials. Currently, RATsApp is limited to VPN users from the RPTU Kaiserslautern-Landau and LMU Munich for security reasons. The system is currently used at these universities and we report results regarding perceived usefulness, user acceptance and usability.

In this manuscript, we will first provide a comprehensive overview of the theoretical background in Section~\ref{background}. Following this, we explore the methods used to evaluate RATsApp in Section~\ref{methods} and describe its capabilities in Section~\ref{functions_n_features}. In Section~\ref{Usability}, we discuss the initial results of our evaluation. 
Finally, we explore in depth our findings and their implications in Section~\ref{discussion}. We then conclude our paper in Section~\ref{conclusion} with a comprehensive summary of our research.

\section{Theoretical background}\label{background}

\subsection{Formative Assessments}\label{formative_assesments}
{\em Formative assessments} are used in the general classroom setting to diagnose students’ difficulties and level of understanding~\citep{shute2008focus, bennett2011formative}. These assessments are not intended to grade or rank students, but rather to provide feedback that can help them improve their understanding and skills. Such evaluations are important for educators to adapt their teaching methods and for students to improve their knowledge through formative feedback (see Section~\ref{feedback_theory}). 

\subsection{Formative Feedback}\label{feedback_theory}
 {\em Formative feedback} is feedback that occurs during a learning process \citep{narciss2008feedback}. 
In general, formative feedback is known to play an important role in improving learners' understanding and enhancing
their skills \citep{butler1995feedback}. Additionally, formative feedback activates learners' metacognitive processes, such as self-improvement and self-monitoring \citep{butler1995feedback}. Hence, it is important to provide computer-based learning environments that are capable of providing formative feedback.

For formative feedback to be most effective, its content should answer the following three questions~\citep{hattie2007power, wisniewski2020power}: (1) Where am I going? (What content do I need to know?), (2) How should I proceed? (How much of the content do I know in relation to what I need to know?) and, (3) Where do I go now? (Where do I go next for more information?). Most importantly, \cite{hattie2007power}, and later \cite{wisniewski2020power}, argue that there are four different levels at which effective feedback can be given: the task level, the process level, the self-regulatory level, and the self level. Feedback on the latter level has been shown to have little or negative impact on students~\citep{wilkinson1980relationship,kluger1998feedback,wulf1979informational}, so it will not be taken into account here. Task-level feedback includes information on the performance in a task, such as correct/incorrect. Process-level feedback is intended to fill gaps in understanding of strategies for completing the task and for detecting errors. Self-regulatory feedback is internal feedback from learners to monitor and regulate their learning. This is important for building confidence and developing self-efficacy.

Additionally, it is important to note that the effective utilization of feedback, and thus RATsApp itself, is contingent upon the student’s ability to accurately interpret and apply the information provided. Feedback, regardless of its appropriateness or comprehensiveness, can only have a meaningful impact if it is understood and acted upon by the recipient~\citep{ hattie1998assessment}.

One type of formative task-level feedback we use is known as {\em elaborate feedback} \citep{kulhavy1989feedback}. This type of feedback not only indicates whether an answer is correct or incorrect but also provides more detailed information on why the answer was correct or incorrect. This feedback is provided by experts in a specific and elaborated manner \citep{kulhavy1989feedback}. It contains an {\em evaluative component} (correct/incorrect information) along with an {\em informative component} that includes information related to the topic and specific details about the task, such as why the given answer was incorrect and guidance for students to arrive at the correct solution. Elaborate feedback has already been reported to be more effective than simply stating whether the answer is correct or incorrect \citep{bangert1991instructional,pridemore1995control}.

The formative feedback mentioned above can be generated through two primary methods: expert-driven and data-driven. {\em Expert-driven feedback} is provided by individuals with a high level of knowledge or skill in a particular area~\citep{deeva2021review}. {\em Data-driven feedback}, in contrast, is generated through the analysis of data~\citep{deeva2021review}. The data analyzed in this way can be presented in a user-friendly manner, for example, through the use of dashboards.

\subsection{Scaffolding}\label{Scaffolding}
Scaffolding is an instructional technique that involves providing learners with support and tools to facilitate engagement with a task or problem~\citep{puntambekar2022distributed}. These tools, known as scaffolds, can take various forms, including hints \citep{pea2004social}, prompts \citep{bannert2013scaffolding}, and explanations \citep{quintana2004scaffolding}. By offering an appropriate level of support at the right time, scaffolding can facilitate the gradual development of learners’ understanding and skills~\citep{hartman2001metacognition,lin2012review}. 

In education, scaffolding is used to fill cognitive gaps, which are the differences between a student's current knowledge and the domain knowledge expected at a certain point during their education. For example, if a student struggles to understand a course text due to their reading level, the teacher may use a didactic scaffolding approach. This approach gradually improves the student's reading skills until they can independently understand the assigned text without help~\citep{van2015effects}.

Through explanations and guidance, students can develop the ability to effectively evaluate the accuracy of their responses~\citep{wu2012pedagogical}. In this way, scaffolding not only supports students in accomplishing tasks that might be beyond their individual capabilities, but it also cultivates new abilities and insights that empower them to handle similar tasks on their own in the future~\citep{van2015effects}.

\subsection{STEM competencies}\label{competencies_theory}

One of RATsApp’s objectives is to facilitate STEM learning by evaluating the following three competencies: 
\begin{enumerate}
    \item {\em Mathematical literacy} involves problem-solving skills, such as using mathematics to solve real-world problems by translating them into a mathematical expression, interpreting results, making predictions, and making reasonable judgments based on these results \citep{pisa2022mathematics}. For instance, a correlation between physics students' mathematical skills and their exam grades is well documented
\citep{blumenthal1961multiple, hudson1982combined, korpershoek2015relation}. Additionally, mathematical ability shows the strongest correlation with learning gain compared to prior knowledge and spatial ability in physics courses \citep{hake2002relationship}. This suggests that students with poor mathematical understanding may struggle in future physics classes. Similar results have been obtained for other disciplines \citep{ballard2004basic,pozo2006requiring,llamas2012mathematical}.

\item {\em Data literacy} is defined as the combination of  efficient behaviors and attitudes that enable individuals to effectively carry out all the steps involved in creating value or making decisions from data \citep{schuller2019future}. It is a foundation for data-based decision making \citep{kippers2018data} and a crucial competence in the 21st century, as it enables a person to analyze and interpret data, to provide others with relevant data, and it facilitates the systematic generation of knowledge based on data \citep{schuller2019future}.

Data can be presented in various forms, such as tables or graphs, and the 
understanding of graphs is a prerequisite for learning in most subjects in higher education \citep{bowen1998lecturing}.

\item {\em Representational competence} involves the ability to extract information from representations, such as graphs, tables, diagrams or picture, to translate between different types of representations, and to interpret and construct representations \citep{kuchemann2021inventory}. Research has shown that students’ problem-solving performance is strongly influenced by their representational competence \citep{kohl2006effect, edelsbrunner2023relation}.
\end{enumerate}

These three competencies are therefore essential in STEM learning. A first step to increase the  respective competence levels is by offering learners feedback on their current competence levels, based on their performance. 

\subsection{Learning Dashboards}

Learning dashboards serve as a tool to facilitate learners’ comprehension of their progress and performance through the provision of easily interpretable visualizations of their data~\citep{verbert2013learning, verbert2014learning,duval2012learning}. For instance, if a student wants to check their progress for the week, they could select a ``Progress" tab on the dashboard. This action would reveal a chart that illustrates their accomplishments for each day of that week. These dashboards capture and display traces of learning activities with the aim of promoting awareness, reflection, and sense-making~\citep{duval2012learning}. 

Additionally, they enable learners to establish goals and monitor their progress towards achieving them~\citep{verbert2014learning}. A student could, for example, establish a target to enhance their rate of learning in the upcoming month, and utilize the dashboard to monitor their daily advancement towards this target. Research has demonstrated the utility of dashboards in educational contexts. For instance, \cite{toyokawa2023active} found that students’ dashboards were effective in facilitating foreign language reading.

Additionally, dashboards may contain data on students' performances to support teachers in diagnosing student difficulties~\citep{knoop2021equalizing}. They allow teachers to rapidly identify areas in which students may require additional support and adjust their pedagogical approach or select an appropriate learning activity accordingly~\citep{van2014supporting}. 
For instance, a teacher might choose to integrate more activities focused on the specific subject into the teaching plan, after observing that a student regularly performs poorly on quizzes.

\subsection{Ethical Considerations}

It is important to consider ethical aspects during the development of educational feedback systems. However, at the moment, this is still rare. In this section, we briefly introduce the ethical guidance for RATsApp.

Prior to the development, we reviewed existing ethical guidelines for the development of digital technologies and systems \citep{hleg2019ethics, Montreal2018, IEEE2021}. Some of them focus specifically on the teaching and learning context \citep{LearningAIEthics}. We are critical of that insofar as they are often too general or schematic or do not refer to the teaching and learning context (with the exception mentioned above). Furthermore, they all refer primarily to technical aspects, such as robustness, or concentrate on questions concerning data security. These belong rather to the legal than to the ethical sphere. Hence, the underlying image or concept of man, in the philosophical sense, within the practical field of learning and teaching is also disregarded \citep{kemp1992unersetzliche, zichy2017menschenbilder, JoistenKultur}. 
Given this background, the ethical orientation – that enables the ethical development of a product in the first place – is currently missing.

Our ethical approach aims to reflect the practical field of learning and teaching at university in STEM subjects – which are in the introductory phase – with the help of \textit{Technoethics for Emerging Digital Systems} (TEDS) \citep{Joisten2022}. This approach considers the various ethical challenges that may arise in relation to digital technologies, incorporating different ethical modes of access. To put it more precisely, the implementation of a new technology within a specific context, such as a learning environment, can only be effectively guided by diverse perspectives of a complex technological ethics, as exemplified by TEDS. Areas of TEDS enable this ethically appropriate perspective concerning: \textit{Anthropo}-, \textit{Pre}-, \textit{Meta}-, \textit{Principle}-, and \textit{Applied}-Technoethics.

These approaches can be summarized as follows: the view of humanity or the concept of man, that guides the field of practice, is fundamental to Anthropo-Technoethics. If we pay attention to the practical field of learning and teaching, it can be noted that humans have the capability to learn in order to become (and this is a normative assumption) more autonomous and to improve their ability to think critically. If this capability is developed with the help of lecturers and peers, it becomes an inherent capability that can also be integrated into a cultural and social context under favorable external circumstances \citep{nussbaum2009creating, sen1997resources, sen1987ethics}. From the perspective of Anthropo-Technoethics, this implies that the extent of the impact of the technical product on learners must be determined and analyzed in terms of its impact on the human being. The guiding question is whether or not the system facilitates or impedes individual autonomy, as well as the users' motivation and interest. However, successful learning is often perceived as the mere acquisition of knowledge. As a result, the effectiveness of digital learning systems is mostly assessed in terms of this knowledge acquisition without considering other abilities and competencies that are also important to  associate with it. Anthropo-Technoethics addresses this gap.

The pre-technoethical perspective focuses on the technical product, in our case the feedback system, which is based on short tests. This approach aims to uncover any potential ethical concerns while the development of the system itself is still in progress, as digital technologies may already harbor ethical problems. The issue of inadequate reflection is evident, for instance, in computer programs such as COMPAS\footnote{\url{https://www.propublica.org/article/machine-bias-risk-assessments-in-criminal-sentencing}}, which predict the likelihood of recidivism or risk rates for offenders in the United States and exhibits discriminatory practices towards people of color. By identifying and analyzing these ethical questions within an interdisciplinary team, digital technologies can be technically adapted and enhanced. Considering the current state of the project, such technical issues are not identified.

Meta-Technoethics pursues theoretical considerations at a more abstract level. For example, inappropriate connotations can be exposed and problematized here by analyzing the use of terms for digital technology. The objective of this perspective is not only to deconstruct (self-evident) concept transfers of human qualities to digital systems, but also to find more appropriate terms. An example, in the context of AI, is the expression `decision-making'. This is an anthropomorphism, since digital systems do not make decisions in the ethical, philosophical sense of the word, but present calculated probabilities \citep{Kaminski2020}. In the current state of the project, category errors are not given in the meta-technical sense.
Applied-Technoethics is mostly concerned with technology assessment. One of its main tasks is to analyze the potential and effects of scientific as well as technological developments comprehensively and with foresight, and to explore the associated social, economic, ecological opportunities and risks, as described in the archive of the German Bundestag\footnote{\url{https://www.bundestag.de/webarchiv/Ausschuesse/ausschuesse19/a18_bildung/technikfolgenabschaetzung}}. In our project, we focus on technology assessment. It is the central task of Applied-Technoethics to critically examine the respective digital application within a concrete context. For example, it is conceivable that the impact on students working with RatsApp reveals unforeseen anomalies that have to lead to a correction of the application. This implies that Applied-Technoethics principally centers on the assessment of technology to ultimately facilitate modifications to the application. In our project Principle-Technoethics is an important foundation. Principles are the normative basic orientations in the learning context.
They can be defined as basic values, basic beliefs, and targets (to be realized) that guide the action in a field of practice \citep{beauchamp2019principles}. 
In the teaching and learning context, the following \textit{primary principles} related to learners can be identified in a topical method, among others \citep{JOISTEN+2004+541+552, Joisten+2006+53+62}: development of autonomy, problem-solving behavior, independence, voluntariness, (self-)responsibility, learning success, learning how to learn, cooperative learning (alternately helping and receiving help), and creativity. These must be taken into account for optimal learning in the interaction between students, lecturers, and RATsApp. By orienting towards these principles, the extent of learners’ alignment becomes clear, which eventually leads to an attitude that enables them to increase their critical self-reflection \citep{JoistenEthikuDigitalisierug}. In the sense of a Principle-Technoethics approach, the best possible education for humans and thus ethical action – with the assistance of new technologies – can succeed only with an appropriate consideration of these principles.
We aim to take a first step in this direction with RATsApp.

\subsection{Research Aims}

Given the pressing need for cost-effective educational learning systems that can also support large-scale research, our study is centered on investigating the application of open-source systems like RATsApp within the context of university STEM education. Our primary focus is to gauge students’ reception of these systems.

In order to assess the usability of RATsApp, we use a theoretical extension of the technology acceptance model developed by~\cite{venkatesh2000theoretical}. This approach will enable us to analyze and interpret survey data from students, thereby providing insights into crucial factors such as the system’s perceived usefulness, its ease of use, and students’ behavioral intentions towards using it.

\section{Materials and Methods}\label{methods}
\subsection{Participants}\label{participants}

The system was implemented as a stand-alone version in the winter semester 2022/23 at two German universities: the RPTU Kaiserslautern-Landau and the LMU Munich. At the RPTU, it was tested in four lectures: Higher Mathematics for Civil Engineers I, Fundamentals of Mathematics II for prospective mathematics teachers, Fundamentals and Applications of Probability Theory and Mathematics for Economists. At the LMU, RATsApp was tested in three first-semester lectures: \textit{Experimental Physics 1}, \textit{Experimental Physics for Veterinary Students} and \textit{Mathematical Methods in Theoretical Physics}.
A total of 64 people who attended these lectures participated in a survey to evaluate usability, as described in Section~\ref{Usability_methods}. The students' participation was voluntary. As additional benefit of using the system besides the learning support, students who actively used the app took part in a lottery of 15 times 20 Euros in each lecture. This lottery took place three times during the semester. Actively using the app meant participating in a cross-lecture questionnaire, which will be discussed later in Section~\ref{students}, solving RATs during the initial four weeks of lectures, solving RATs throughout lecture weeks 5 to 10, solving RATs from the 11th week of lectures as well as course completion. The course completion was determined by either reporting the final exam grade or opting out with a valid reason. The lotteries were based on participation only and did not require fast or correct responses. However, we imposed a minimum of four separate logins with at least 24 hours between them to qualify for the raffles that involved solving RATs.

\subsection {Usability evaluation}\label{Usability_methods}	
To measure how the students interact with the system, we created a usability questionnaire. From the extension of the technology acceptance model (TAM2)~\citep{venkatesh2000theoretical}, we used the sections \say{Intention of use}, \say{Perceived Ease of Use}, \say{Perceived Usefulness}, \say{Job Relevance} and \say{Output Quality}.
We translated these sections to German and adapted them to the context of RATsApp. For example, we renamed the \say{Job Relevance} section to \say{Relevance to the Course of Study}.
Additionally, we added questions asking for age, gender, frequency of use, attended lecture, and if the participants have an RATsApp account. Furthermore, we added free text fields after each section leaving room for additional feedback.
The questionnaire we used, as well as an English translation of it, can be found in the Appendix~\ref{secA1}.
The survey was conducted anonymously online and participation was voluntary. 
More details about the lectures that took part in the survey can be found in Section~\ref{participants}.

\section{Functions and Features of RATsApp}\label{functions_n_features}

\begin{figure*}
    \centering
    \includegraphics[width=\linewidth]{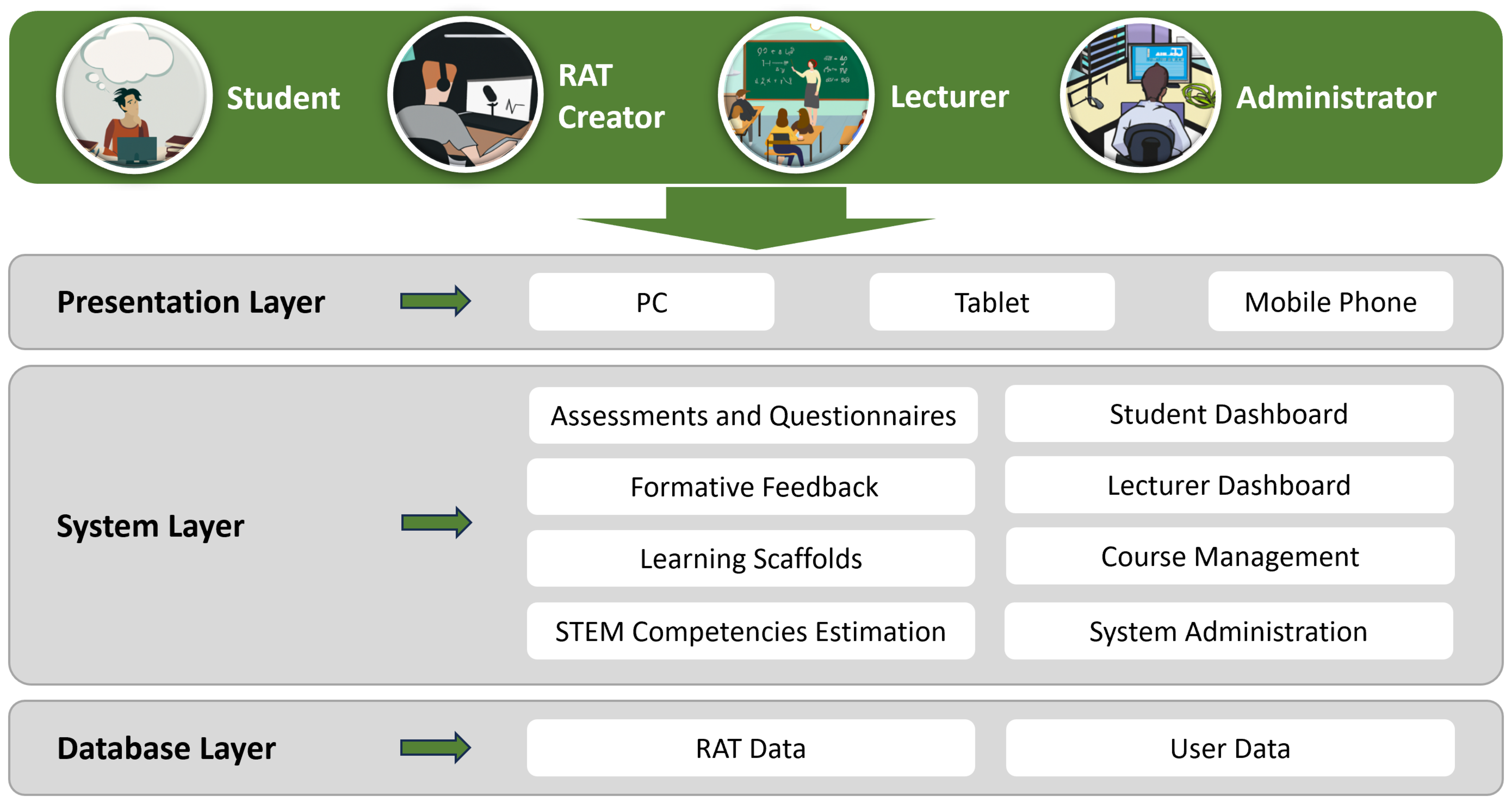}
    \caption{RATsApp supports four distinct user types: Student, RAT Creator, Lecturer, and Administrator. The architecture of RATsApp is divided into three layers: the Presentation Layer, the System Layer, and the Database Layer. Each layer serves a specific function in the overall operation of the application.}
    \label{fig:system}
\end{figure*}

The system has been designed to accommodate four distinct user types: student, RAT creator, lecturer, and administrator. The specific capabilities of each user type will be elaborated upon in Section~\ref{users}. In accordance with the architectural design proposed by~\cite{liu2022towards}, the system is composed of three layers, as shown in Fig.~\ref{fig:system}: the presentation layer, which handles the user interface; the system layer, which manages the application's logic and functionality; and the database layer, which stores and retrieves data. 

\subsection{The Presentation Layer}
The presentation layer of the application is responsible for rendering the user interface. To enhance the user experience, front-end technologies such as HTML, CSS, Jinja2, and JavaScript have been employed. The application has been designed to be responsive and scale appropriately across a range of devices, including computers, tablets, and mobile phones. Additionally, the interface style was tested across the most commonly used web browsers to ensure a consistent and positive user experience.

\subsection{The System Layer}

\subsubsection{RATsApp Assessments and Questionnaires}\label{Assesments}
In this system we test the abilities of students by using assessments called RATs, which are short and quick formative assessments (see Section~\ref{formative_assesments}). RATs are created using the recommended materials of the lectures they are intended for, as well as other sources such as books and the internet. The creator of the RAT assigns competence criteria, scaffolds, topics, and concepts to it. Before it is made available to students, two experts must approve the RAT using the system to ensure its correctness.

RATs can be utilized to generate exercise sheets, questionnaires, tests, and cross-lecture questionnaires. RATs that belong to the cross-lecture questionnaires can be used to test knowledge all students benefit from, regardless of their course of study or attended lectures. For instance, these RATs may be used to evaluate various high school proficiencies of first-year students or as a post-test upon completion of a course.

\subsubsection{RATsApp Feedback}\label{feedback}

RATsApp aims to provide formative feedback that is useful for students to narrow the gaps between their understanding and the curriculum. To enhance the learning experience, the system provides constructive feedback as students complete each RAT. 
The feedback provided by the system is informed by expert knowledge and is delivered through the use of elaborated feedback techniques, as outlined in Section~\ref{feedback_theory}. This feedback is provided immediately following the completion of a task and offers students detailed insights into their performance. The informative component of this feedback could include the following information: i) an explanation to guide the student in identifying why the chosen answer is incorrect, ii) information to guide the student through steps on how to solve the task correctly, iii) information to guide the student to external scaffolds, or iv) information that includes a combination of all or some of the above options. It is up to the expert to decide what the most relevant piece of information to give is and what kind of feedback is needed according to the task. Therefore, this feedback helps students at the task level and/or the process level, depending on the difficulty of the task.

\subsubsection{RATsApp scaffolds}\label{Scaffolding_methods}
During the task, students are offered optional scaffolding support (refer to Section~\ref{Scaffolding}). This support is not automatically provided through pop-ups or similar means; rather, students must actively access it by clicking on a tab adjacent to the task. The scaffolds are related to the topic and concept of the task. It can be provided in the form of a text, a link to external online material or a reference to learning material. We argue that this helps students who are at a self-regulatory level, i.e., they are able to control their learning process (see Section~\ref{feedback_theory}). 

From an ethical perspective, it is important to emphasize that although an active decision must be made to access the provided learning materials, it is still essential to consider their ease of access. Thus, students are not facing an experience of resistance~\citep{wiegerling2021exposition}. On the other hand, such a resistance may be very positive in the learning process, as induced through the independent overcoming of a resistance (here e. g. conducting their own research for suitable materials), students do gain the attitude that they can attribute the merit of self-learning – despite some difficulties – to themselves. This increases their problem-solving skills as well as it deepens their self-confidence \citep{fuchs2020verteidigung}. Nonetheless, the provided – easily accessible – material can still support students in the learning process to distinguish verified from unverified sources through the indicated websites, books, and links.

\subsubsection{Skill level estimation of STEM competencies}\label{competencies}

We aim to estimate the three key STEM competencies: mathematical literacy, data literacy, and representational competence. The main idea is to estimate which skill level is needed to solve a RAT and then to deduce from the correctness of the solved RATs the overall competence level of the student relative to the skill level needed for all already solved RATs.

To estimate the competencies defined in Section~\ref{competencies_theory}, we created a list of objective criteria for RATs. All these criteria are connected with one or more competence and can be used to estimate which level of skill per competence category is needed for a RAT.
Our list was based on the criteria described by Scheid and colleagues \citep{scheid2017erhebung}. Furthermore, we used the definition and criteria of the mathematical literacy of the Programme for International Student Assessment (PISA) 2022 mathematics framework \citep{pisa2022mathematics} and the definition and criteria of a framework for data literacy \citep{schuller2019future}. In order to make a list suitable for evaluating a large number of RATs, we have compressed the information into just a few list items. As a starting point we subjectively decided which criteria of our sources to select and which we wanted to combine into one criterion. In this way, we obtained a list consisting of 21 items (see Appendix~\ref{secA}). Please note that the list of questions needs to be validated in further studies. Apart from that, the competencies are not always clearly separable and therefore some criteria belong to more than one competence.

To estimate the relative competence levels of the student, the system counts the number of competence category criteria fulfilled by a RAT. This number is added to the number of points the student could have reached at maximum for this competence. If the RAT is solved correctly, it is also added to the current score. By dividing the current score by the maximum possible points, we obtain a relative score for the competence. A graph visualizes the connection between competence level and number of solved RATs for each competence  after every completed RAT. This feedback is intended to help to identify weaknesses and strengths in one or more of these competences. This is for the purpose of guiding the students on what kind of tasks they should work on more to improve the final score in their lesson. 
In the future, a machine learning (ML) algorithm could be trained to automatically assign those RATs to students that could help them improve a specific competence.

\subsubsection{Student Dashboard}\label{student_dashboard}
A user-friendly dashboard displaying key performance metrics is also available to students for tracking their progress. The task performance statistics provided in the dashboard include the percentage of tasks answered correctly or incorrectly and the total number of RATs answered per week. This data-driven feedback aims to help students regulate their learning process and activate the necessary meta-cognitive processes necessary for self-regulation.

From an ethical and didactic standpoint, it is crucial to consider the potential for demotivation among students when confronted with transparent data regarding their individual task completion statistics~\citep{hattie2008visible}. As already mentioned (see Section~\ref{feedback_theory}), a certain level of competence in interpreting and handling feedback is required.

\subsubsection{Lecturer Dashboard}\label{lecturer_dashboard}
Lecturers have access to dashboards that display students aggregated statistics and difficulties. They can monitor their students’ performance on RATs and individual competencies. This feature allows lecturers to conduct formative assessments and evaluate the effectiveness of their teaching methods. The dashboard also displays tasks that are frequently answered incorrectly, which are displayed in the following error categories:

\begin{itemize}
    \item Always answered incorrectly.
    \item Often answered incorrectly (answered correctly less then 40\% of the time).
    \item Deceptive answer (the same wrong answer was selected more then 30\% of the time the question was answered).
\end{itemize}
In addition, lecturers can see the most frequently chosen answer option. This can help them identify misconceptions their students may have and intervene.
This feedback can contribute to reconsidering one's responsibility as a lecturer and the teaching methods in a self-reflective way, selecting new approaches, and ultimately implementing them.

\subsubsection{System Administration}\label{admins}
A system administrator is responsible for granting access rights to new users. Upon request, an admin can also delete an account while retaining non-personal data and log entries.
Furthermore, admins also receive data-driven feedback from the system. They have access to statistics for all available lectures. In addition, admins can see how many RATs are generated per lecture per week and which user ID created them. Also, admins can see how the users interact with the site by monitoring log entries.

\subsubsection{Course Management}\label{course_managment}
This component of the system layer is responsible for managing all course-related content. This includes the creation of RATs, the generation of weekly RAT sheets, and the development of lecture materials, among other tasks. These features are further explained in Section~\ref{functions_n_features}. The system is designed to facilitate the efficient organization and delivery of course content to enhance the learning experience for students.

\subsection{Database Layer}\label{secure_python_application}

The database architecture of RATsApp is divided into two distinct components: the RAT database and the user database. The RAT database stores the data required for the application's operation, while personal user data is stored separately in the user database to mitigate the risk of inadvertent exposure due to programming errors. Security is a paramount concern for the system, and as such, widely used and well-documented frameworks are employed for server-side code development and database access. 
RATsApp is written as a Python web application using flask\footnote{\url{https://flask.palletsprojects.com}} and SQLAlchemy\footnote{\url{https://www.sqlalchemy.org/}}. It runs on Gunicorn\footnote{\url{{https://gunicorn.org/}}} as a wsgi server and is served from the web using nginx\footnote{\url{https://nginx.org/}} as a reverse proxy. 

These frameworks are actively maintained and provide inherent protection against common security vulnerabilities, including SQL injection attacks \citep{lerner2013forge}. 
MariaDB\footnote{\url{https://mariadb.org/}} is used as an SQL database to save the data. Furthermore, the server is only accessible from the networks of the universities of LMU Munich and RPTU Kaiserslautern-Landau as well as from the corresponding VPN connections. This is meant to lower the number of potential intrusions from outside.

\subsection{Users}\label{users}
When registering, RATsApp only requires a valid email (belonging to one of the institutions from which registration is accepted) and a password. In addition, the user must accept the terms of use and choose one of the following four user categories: students, RAT creators, lecturers or administrators, with user access rights increasing in this order. This means for example that RAT creators have access to all the content that students have access to, and so on and so forth, unless otherwise specified. Below we will explain the different categories of users and what content they can access. We start with users with a low number of access rights (Students) and continue in ascending order. To differentiate between different user types and keep track of the interaction history we created a sophisticated account system that allows them to access features commonly found on websites, like email validation, signup, or changing the password. It is possible to customize the own profile by selecting which statistics should be displayed.

\subsubsection{Students}\label{students}

\begin{figure}
    \centering
    \includegraphics[width=\linewidth]{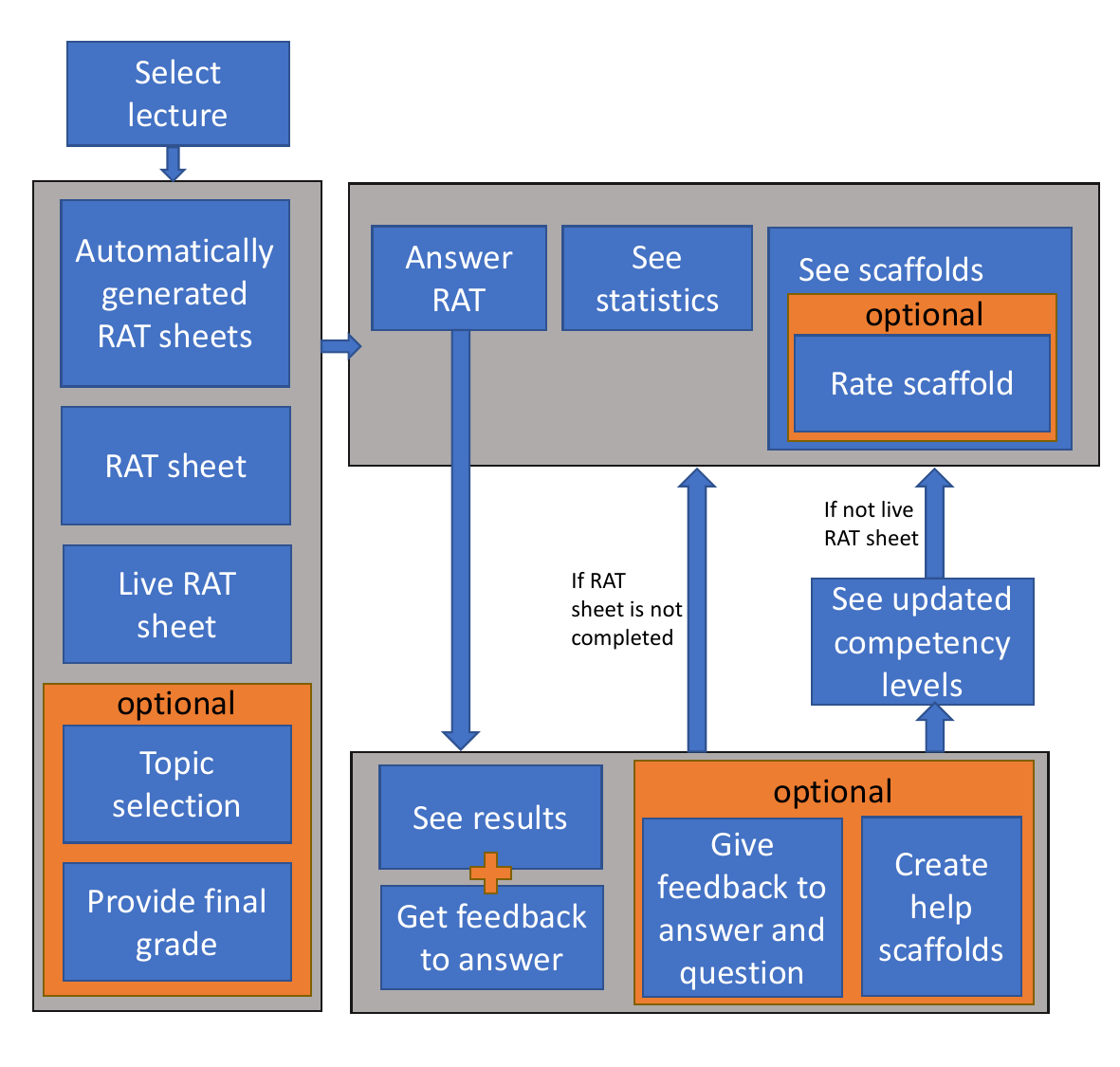}
    \caption{\emph{The student workflow of answering RATs:} First a lecture has to be selected, afterwards a student can choose either one of answering automatically generated RATs, a RAT sheet, or a live RAT sheet. This will lead to a new page with the choice to answer the RAT, view scaffolds or statistics. After answering a RAT results and feedback are shown and the student is either guided to the next RAT in the sheet or shown the now updated competence levels, if the sheet is finished.}
    \label{fig:student_flowchart}
\end{figure}

Students can register for a lecture by first searching for the lecture in the systems course catalog and then entering the lectures password, which is usually provided by the instructor. In their profile, students have access to a sidebar in which they can select the lecture to work on  (see Fig.~\ref{fig:student_flowchart}) or if a cross-lecture questionnaire is available to them. The cross-lecture questionnaire is an option only available to students and not for other user types. It can be used to test fundamental knowledge that all students benefit from, independent of the lectures they attend.

When a lecture is selected, students can view the RATs that were either automatically selected or selected by the lecturer manually, RAT sheets or live RATs. The latter are only accessible when a live session is in progress. More details about these options can be found in Section~\ref{lecture}.
The topics of the RATs can also be limited by a topic selection function. A student accessing any of the above options, will be presented with the first available RAT. In addition, at the top of the RAT a tab bar will appear that allows the student to navigate to the overall statistics of the answered RATs or to display and, optionally, rate scaffolds. As explained in Section~\ref{student_dashboard}, the student statistics presented in the tab include the percentage of RATs answered correctly/incorrectly and the total number of RATs answered per week. In addition, as explained in Section~\ref{Scaffolding_methods}, the scaffolds provided in this tab can be suggested feedback from experts in the form of text, a link to a video, a link to external online material, or a reference to a book. This suggested material is related to the topic and concept of the RAT that the students are working on.

After answering a RAT, the student receives elaborate feedback created by an expert (see Section~\ref{feedback}). 
 Furthermore, a students also has the option to suggest scaffolds that were found useful in answering the question, leave comments and to point out possible errors in the RAT.  Afterwards, the student moves on to the next RAT or, if the RATs were completed, is shown the newly updated competence  levels presented in Section~\ref{competencies}. This is a way to determine if a student is having difficulties with any of the skills associated with specific competencies. The complete student workflow is visualized in Fig.~\ref{fig:student_flowchart}.

\subsubsection{RAT creators}
A RAT creator can be a teaching assistant, an administrator, or a lecturer. All RATs can contain the following information:

\begin{enumerate}
\item {\bf Subject/Topic:} A subdivision of the lecture such as “Matrix operations”
in a Mathematics lecture. With this thematic category of the RAT, they can be searched in the database. Hence, the RAT can be used in different lectures. In addition, this category specifies the knowledge required to solve the RAT correctly.
\item {\bf Concept:} A concept is a more specific label within the subject, such as, in the case of the subject ``Matrix operations", a related concept could be ``matrix multiplication". In other words, concepts are a more fine grained subdivision of a topic. 
\item {\bf Lecture:} At the moment of creation, the RAT can be assigned to one or more lectures which enables it to occur in automatically generated questionnaires if all connected topics and concepts have been taught in the lecture.
\item {\bf Type of question:} The RAT creator can select three types of questions; multiple choice, multiple True/False, or open-ended.
\item {\bf Answer specific or elaborated} feedback: When creating the right/wrong answers of the multiple choice, the RAT creator can add individual feedback to the specific answer. As explained in Sections~\ref{feedback} and \ref{students}, this is with the purpose of helping the students to understand why a certain option was the incorrect/correct one.   
\item {\bf Competence category:} The RAT creator can label it by choosing options from a pre-determined list (see Appendix~\ref{secA}) that helps to estimate how much of a specific competence is needed to correctly solve the RAT. The considered competencies are mathematical literacy, data literacy and representational competence.
\item {\bf Scaffolds (Optional):} As explained in Sections~\ref{Scaffolding_methods} and \ref{students}, scaffolds can be given to the students related to the subject/topic and concept of the RAT. These scaffolds can be in the form of text, links to videos, links to external online material, or book references. Students can view the scaffolds while solving a RAT.  
Scaffolds differs from the other items in the list since it is optional and can be added to a RAT later.
\item In addition, the creator of the RAT may select an additional option, by assigning it to be part of a {\bf cross-lecture questionnaire} (see Section~\ref{Assesments}). If a RAT is marked as belonging to a cross-lecture questionnaire it can not occur in regular RAT sheets or as an automatically selected RAT.
\end{enumerate}

A powerful embedded text editor (Quill\footnote{\url{https://quilljs.com/}}) is used to create the RAT question, the answers and the scaffolds. This editor supports 
latex and images.
Once a RAT is created, it can be edited, duplicated, deleted, revised (see below), and searched by author, lecture, topic, or concept.  
Since the quality of a RAT is of great importance, it is not immediately available to students at the time of its creation. This is the case even when a RAT is linked to a lecture in the database. To maintain quality control of the RAT, a review function was implemented. In this option, at least two experts in the field (this number can be changed in the configurations) of the corresponding RAT have to approve the item together with the corresponding feedback so that it can be administered in the lecture. The reviewer has the opportunity to communicate with the creator of the RAT through a comment function  
in order to clarify any doubts or possible errors. The RAT creator will receive a notification email when comments are made on the RAT. The email includes the RAT id and the comment itself. Anyone who makes a comment will receive an email notification of new comments. This reviewing process ensures that RATs used in the automated system (explained in the next Section) exhibit a good quality. 

The same applies to the scaffolds. Its content must be approved before it is made available to the students. 
However, it only has to be approved by one reviewer (this number can be changed in the configurations).   

\subsubsection{Lecturers}\label{lecture}

A lecture can be created by a lecturer or an administrator. 
During the creation of the lecture, the following information should be entered: the name of the lecture, the intended audience, appointment dates, lecture term, and a lecture code or password for students to join. Additionally, a syllabus can be created using the appointment dates for the lecture by adding subjects and concepts to be covered on specific date. This is necessary to automatically adding questions to a test because the system needs to verify if the required concepts to solve a RAT have already been taught.

In order to assign RATs to the students (see Fig.~\ref{fig:student_flowchart}) the lecturer has the following options: 
\begin{enumerate}

\item The lecturers can use the syllabus to {\bf automatically} generate a RAT sheet, which is a set of RATs  related to the subjects and concepts filled in the syllabus. The RAT sheets can be asked in the students profile starting from the assigned date. In this option, the lecturer also has the opportunity to see the pool of possible RATs and remove undesired RATs to the sheet.  \item An additional function for adding RATs {\bf manually} is to select them from a list of RATs belonging to the specific lecture. The created RATs are displayed in the same place as those automatically selected by the system.
\item The lecturers can also create {\bf RAT sheets}. It is created with a unique name and lecturers have the option to select the RATs contained within from the pool of created RATs and give them a desired order of appearance.
\item Lecturers also have the option to use the website for {\bf live RAT sheets}. This lets them select an already created RAT sheet and open a live session where all students in the lecture are able to answer the RATs simultaneously. Statistics describing the students performance for each individual RAT and the whole RAT sheet are shown in real time to the lecturers. 
This feature enables active learning, an teaching method that can increase interaction between students and lecturers, as well as between students themselves, and thereby provides formative feedback in real time~\citep{freeman2014active}.

\end{enumerate} 

As mentioned in Section~\ref{lecturer_dashboard}, lecturers have also access to the statistics of students in a lecturer dashboard.

\subsubsection{Administrators}
An administrator has full access to all areas of the website, including the additional capabilities outlined in Section~\ref{admins}. These capabilities include managing user accounts, moderating content, and configuring site settings, among others.

\section{Results: Usability evaluation}\label{Usability}
To evaluate the attitude of students towards the tool and its output quality, we created a usability survey which was conducted at the end of the winter semester 2022/23 (see Section~\ref{Usability_methods}). Participation was voluntary and there was no need to log into RATsApp for it (see Section~\ref{participants}). Students used either a QR code or a link provided by us to access the test, making it completely anonymous. The only personal information asked for in the survey was optional and consisted of age and gender. In order to further categorize the users, the test included a field to indicate a lecture they attended and whether or not they had created an account in RATsApp. We categorize students as non-users if they stated that they never used the website even if they were registered.
In total, eleven registered users stated that they never used the website and were therefore counted as non-users. Taking this into account, the survey was completed by 36 users and 28 non-users. The average age of the participants is 19.87 with a standard deviation of 3.19 years. Of the participants, 28 were male, 33 were female, and three provided no information.

\begin{figure}
    \centering
    \includegraphics[width=\linewidth]{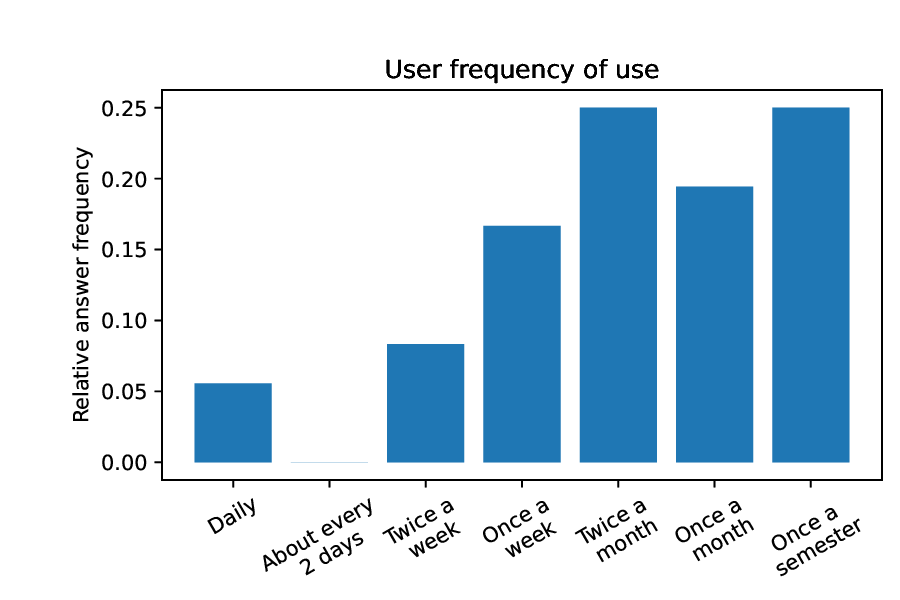}
    \caption{Frequency of all participants which stated that they use the system more often than ``never" and who were therefore categorized as users. The categories have been translated from German to English and their selection frequencies are relative to the number of users.}
    \label{fig:questionnaire_frequency_bar_plot}
\end{figure}

This usability test consisted of three elements: multiple choice questions, a Likert-type questionnaire and written feedback (see Appendix~\ref{secA1}). One of the multiple-choice questions relevant to this survey is related to frequency of use, as shown in Fig.~\ref{fig:questionnaire_frequency_bar_plot}.  The most frequent responses to the usage question were “Twice a month” and “Once per semester,” each accounting for approximately 25\% of total responses. Only around 5\% of respondents reported daily usage. The median usage value corresponds to “Twice a month”.

\begin{figure}[H] 
    \centering
    \includegraphics[width=\linewidth]{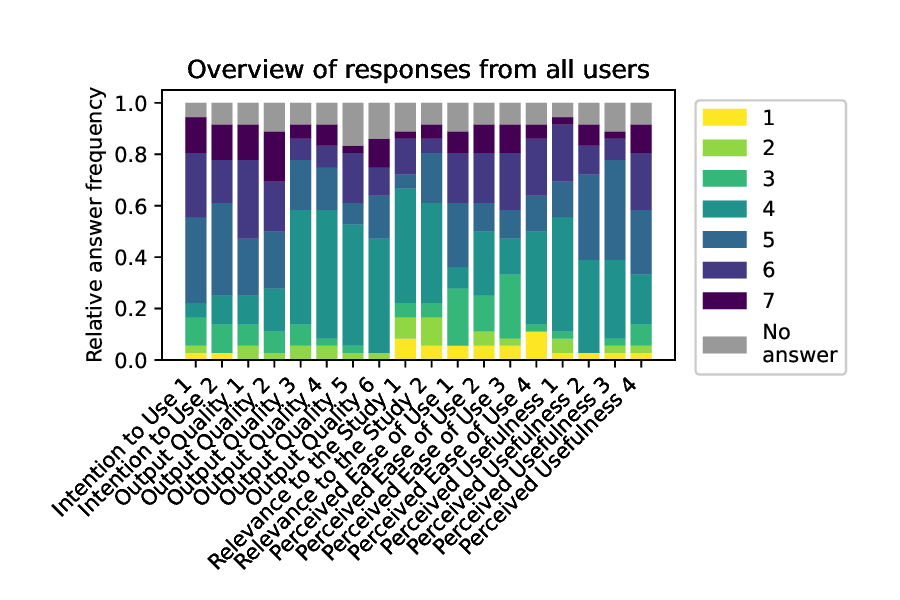}
    \caption{Answer frequencies of use of participants who stated that they use the system more often than ``never" and were therefore categorized as users. The use frequency categories (see Table~\ref{tab:used_questionnaire} for more details), translated from German to English, show the selection frequencies relative to the number of users. The questions belonging to each item can be found in the Appendix~\ref{secA1}.}
    \label{app_fig:questionnaire_frequency_bar_plot}
\end{figure}

\begin{figure}
    \centering
    \includegraphics[width=\linewidth]{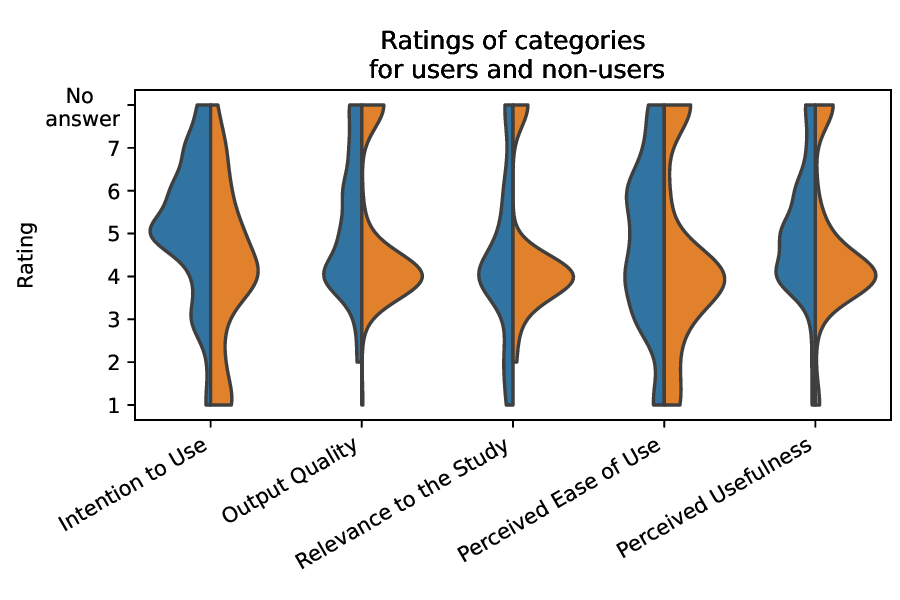}
    \caption{Distribution of selected rating for a category of the questionnaire relative to the number of all participants. The violin plot on the \textbf{left} in each category shows the rating of users of the platform, while the violin plot on the \textbf{right} shows the ratings of non-users. The different numbers in the figure correspond to a scale from 1=``strongly disagree" to 7=``strongly agree" (positive for the platform), with 4=``neutral". The plot also includes a representation of participants who did not provide an answer. The categories were translated from German to English.}
    \label{fig:questionnaire_category_bar_plot}
\end{figure}

The Likert-type questionnaire contained questions of the different categories \say{Intention of use}, \say{Perceived Ease of Use}, \say{Perceived Usefulness}, \say{Job Relevance} and \say{Output Quality}.  Each of the above categories corresponds to the averaged results of a group of several questions listed in Appendix~\ref{secA1}. The responses to each question are depicted in Fig.~\ref{app_fig:questionnaire_frequency_bar_plot}. To facilitate comprehension, we have consolidated the results of all questions according to their respective categories in Fig.~\ref{fig:questionnaire_category_bar_plot}. This figure shows the frequencies of the ratings of these questionnaire categories relative to all participants. The numbers in the figure correspond to a scale from 1=``strongly disagree" to 7=``strongly agree", with 4=``neutral". A rating of 7 on the scale, corresponding to “strongly agree”, indicates a positive outcome for RATsApp. The violin plot on the \textbf{left} in each category shows the rating of users of the platform, while the violin plot on the \textbf{right} shows the ratings of non-users. The plot also includes the frequency of participants who did not provide an answer. In general, most questions were not answered by about 5\% to 10\% of the users. For simplicity we interpret a rating of 5 or higher as confirmation of the question asked, since 4 is the neutral rating.

\begin{figure}
    \centering
    \includegraphics[width=\linewidth]{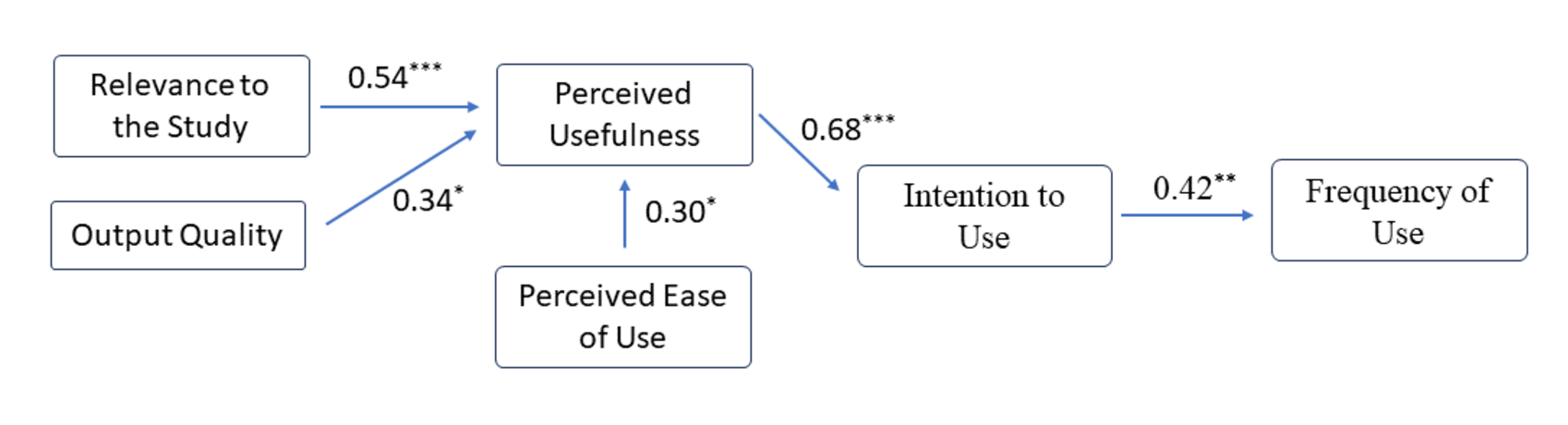}
    \caption{Results of the TAM2 analysis: This figure illustrates the interrelationships between the categories. The significance of these linear relations is indicated as follows: $^*$ denotes $p<$0.05, $^{**}$ denotes $p<$0.01, and $^{***}$ denotes $p<$0.001.}
    \label{app_fig:TAM}
\end{figure}

Most of the questions, except the questions for the intention of use, can only be answered objectively after using the system. Many non-users also answered these questions, often choosing the \say{neutral} answer option. 
For a more detailed evaluation, we focus our subsequent analysis solely on system users. Our findings are derived from individual questions, as illustrated in Fig.~\ref{app_fig:questionnaire_frequency_bar_plot}, which shows the unsummarized answers of the users. Furthermore, we present our results in the context of TAM2 (see Section~\ref{Usability_methods}), as shown in Fig.~\ref{app_fig:TAM}.

{\bf{Intention to use:}} Around 70\% 
of students rated their intention to use higher than the neutral value of 4. Only about 25\% rated it neutral or lower and about 5\% did not answer. 
This is in line with the frequency of use shown in Fig.~\ref{fig:questionnaire_frequency_bar_plot}, where over 25\% of users indicated that they use RATsApp only once in the semester. It should be noted that despite the fact that many users rated their intention to use as 5 or higher, only around 30\% of users use the system weekly or more often. By numbering each answer option for the frequency of use starting with 1 for \say{daily} and ending with 7 for \say{Once a semester}, we find a significant linear relation of 0.42 ($p=0.0014$) between the summarized intention of use (shown in Fig.~\ref{fig:questionnaire_category_bar_plot}) and the frequency of use (shown in Fig.~\ref{fig:questionnaire_frequency_bar_plot}). 
For the non-users, it can be seen that around 35\% have an intention of using the system if they are able to. 

{\bf{Output quality:}} Nearly 70\% 
of users consider the quality of the expert generated feedback to be high, and around 60\% 
of users have no problem with the quality. 
On the other hand, when asked about the quality of the estimation of their competence level, only around 35\% 
of users rated the quality as high and roughly the same amount had no problem with the quality. 
The quality of the display of the temporal progression of the user's answer correctness was also rated by only a bit more than 30\% 
as high but interestingly around 40\% 
had no problem with its quality. 

Our analysis based on TAM2 reveals a significant linear relationship between the output quality and the perceived usefulness ($p$=0.011).

{\bf{Relevance to the study: }}Only around 20\% 
of the students rated the system as important to their studies and 30\% 
as relevant to their studies. Despite these percentages, our analysis based on TAM2 revealed a significant linear relationship between the relevance to the studies and the perceived usefulness ($p=3\cdot10^{-5}$). Interestingly, this relationship was not moderated by the output quality ($p> $0.05). 

{\bf{Perceived ease of use:} }In terms of the perceived ease of use, around 55\% 
of users felt that interacting with RATsApp was clear and understandable. However, only around 40\% 
of users indicated that they could navigate RATsApp without requiring much effort. 
Around 45\% 
of users find RATsApp is easy to use and only around 40\% 
of users find it is easy to make RATsApp do what they want to do.

Furthermore, as proposed by TAM2, we observed a significant linear relationship between the ease of use and the perceived usefulness ($p$=0.027). Contrary to the propositions of the TAM2, we did not observe a significant linear relationship between the ease of use and the intention to use ($p >$ 0.05). Also, when considering the ease of use as the antecedent variable, the perceived usefulness as the mediator, and the intention to use as the consequent variable, we found no significant mediation ($p >$ 0.05). 

{\bf Perceived Usefulness:} It follows that around 40\% 
and 50\% 
of the people think that using RATsApp improves their performance in their studies and that they learn more using RATsApp, respectively. Despite the low percentage of students who found RATsApp relevant to their studies, half of the users 
tend to think that using RATsApp makes them more effective learners and even around 60\% 
of users find RATsApp useful for their studies. 

In line with the propositions of the TAM2, we found a significant linear relationship between the perceived usefulness and the intention to use ($p$=$7\cdot10^{-9}$).

\begin{table}[t] 
\renewcommand{\arraystretch}{1.1}
\caption{Frequencies of the categories of the received feedback\label{tab:feed_back_categories}}
\centering
\begin{tabular}{@{}lp{0.15\linewidth}@{}}
\toprule
\textbf{Category}	& \textbf{Frequency }	\\
\midrule
Cumbersome access to website (VPN, Eduroam, no app) & 19 \\ 
Technical problems/website not working properly & 7\\ 
Missing features in the beginning & 6 \\ 
Cumbersome or unclear navigation of the website & 5\\ 
RATsApp is useful in principle and a good idea & 3 \\ 
Feedback and graphics from RATsApp are helpful & 3\\ 
Participation for monetary reasons & 2 \\ 
No time & 1\\ 
Other & 5\\ 
\botrule
\end{tabular}
\end{table}

Finally, we analyzed the written feedback and created different categories (see Table~\ref{tab:feed_back_categories}). The categories have been translated from German to English and are listed below: 

\begin{enumerate}
\item {\em Cumbersome access to website (VPN, Eduroam, no app):} includes feedback recommending that RATsApp be developed into an application, be accessible from anywhere and that access in general should be less cumbersome. An example of a feedback that belongs to this category is \say{Please get rid of the VPN and introduce an app for RatsApp, if possible also offline}.
\item {\em Technical problems/website not working properly:} contains feedback reporting bugs and other technical problems with the website. Most of these issues are already resolved. An example is \say{Sometimes it did not work as desired (course was not displayed; I could not open RAT sheets on my cell phone)}.
\item {\em Missing features in the beginning:} includes feedback complaining about missing features that were already implemented at the time of the survey, as well as feedback praising  features which were introduced between the deployment of the website and the time of the survey.  For example, feedback that falls into this category might be  \say{The division into the RAT sheets, like on moodle, helped a lot to repeat specific tasks}.
\item {\em Cumbersome or unclear navigation of the website:} contains feedback that does not specify further why the navigation is cumbersome as well as specific feedback. An example for such an feedback is \say{The start page could be easier to use and you could go directly to your questions. In addition, it is not always easy to see the competence level}.
\item {\em RATsApp is useful in principle and a good idea:} contains, for example, feedback that praises RATsApp in general, such as \say{RATsApp is a very good complementary tool to check and improve your learning progress in a playful way}.
\item {\em Feedback and graphics from RATsApp are helpful:} contains feedback that specifically complements the output of our system, such as  \say{Otherwise, the app also shows quite well how the physical skills develop over the course of the exercises and I find the explanations to the tasks immediately following great}. 
\item {\em Participation for monetary reasons:} Some users wrote that they also participated for monetary reasons. So we put feedback like \say{The sweepstakes are really great :) Can use the ``pocket money" well} in this category. 
\item {\em No time:} Only one participant cited lack of time in the feedback. We still put this in a separate category because it was one of our concerns that students would have not enough time for RATsApp because of their other duties. The exact feedback is \say{Unfortunately, I didn't find time to use the app besides working on the practice sheets}.
\item {\em Other:} contains other feedback that was only mentioned once. An example for this is \say{However, the solution paths are not always easy to follow}. 
\end{enumerate}

The frequencies of the free-text categories shown in Table~\ref{tab:feed_back_categories} corroborate  certain findings from the user ratings. The most frequently reported issue was difficulty accessing the website, followed by technical problems and missing features at the start of the semester. These issues likely contributed to the low usage rates of RATsApp despite high reported intention of use and perceived usefulness for studies. Another common feedback was the cumbersome or unclear site navigation, which may explain why less than half of all participants rated the system as easy to use and suggests an area for improvement. Despite these challenges, users expressed appreciation for the concept behind RATsApp and found its feedback helpful.

\section{Discussion}\label{discussion}

In this section, we analyze and interpret the findings from our usability survey, adhering to the structure delineated by the TAM2 framework, as proposed by~\cite{venkatesh2000theoretical}. As mentioned in Section~\ref{Usability_methods}, this framework was also instrumental in the construction of our usability survey.

Our usability survey reveals a strong inclination towards the use of RATsApp. A significant majority of participants, constituting 70\%, expressed an intention to use it. According to the TAM2 framework, the intention to use a system is a key determinant in whether an individual will adopt or reject a new system. In our results, we identified a significant linear relation between the intention to use and the frequency of use. This finding aligns with the principles of TAM2 and suggests that the lack of an intention to use the system could be a primary factor leading to its non-use.
This linear relation highlights the importance of fostering a strong intention to use among potential users. 
Therefore, in our discussion, it is crucial to address any factors that may be impeding this intention in order to increase system usage. 

Regarding perceived ease of use, consistent with TAM2, we identified a significant linear relationship between ease of use and perceived usefulness. 
However, diverging from TAM2, we did not find a significant linear relationship between ease of use and intention to use. In the TAM2 paper, this relationship was found to have a minor effect, leading us to hypothesize that our sample size may not be large enough to detect this relationship. Similarly, unlike TAM2, we did not identify any significant direct or indirect effect in the mediation with ease of use as the antecedent variable, perceived usefulness as the mediator, and intention to use as the consequent variable. This could likely be due to the same reason mentioned above.

In terms of perceived usefulness, our findings reveal a significant linear relation between perceived usefulness and intention to use, which is in line with TAM2. We also observed a significant linear relation between output quality and perceived usefulness, as well as a highly significant linear relation between relevance to the studies and perceived usefulness. These findings are consistent with TAM2. However, unlike TAM2, this linear relationship is not moderated by output quality.

Additional external factors, some of which are considered by the TAM2 but not evaluated in this study, could also contribute to the perceived usefulness. For instance, students might feel compelled to use the app due to external influences or expectations, such as recommendations from teachers or peers, regardless of their personal assessment of its usability or utility. 
As discussed by \cite{pfeffer_organizations_1982}, adapting a certain behavior, like using a system, could enhance a person's standing within the group if influential members of that person’s workplace social group believe that it should be adopted.

In light of our analysis, which did not find a significant linear relation between the ease of use and the intention to use RATsApp, it appears that enhancements in the user interface, such as design and layout modifications, may not significantly increase its usage. A more effective strategy for boosting RATsApp usage could be to increase its relevance to the course of study. This could be achieved, for instance, by better aligning the RATs with lecture content and examination material. It is also crucial that the benefits of the system are effectively communicated to users to emphasize its relevance.

Additionally, our results suggest that while less consequential, refining output quality could also contribute positively to its usage.
The output quality subcategories include some of RATsApp’s most critical features, making their improvement also a priority. One such feature is the assessment of STEM competency, which includes data literacy, mathematical, and representational competencies. The usability test indicated potential for enhancement, with only 35\% of students rating the output quality as high. 

A further aspect associated with output quality is the dashboard. Merely 30\% of students rated the dashboard of RATsApp highly, indicating a potential area for improvement. The visualization of the graphs in the dashboards may require enhancement to improve readability and comprehension, a critical factor considering that learning dashboards are known to promote awareness, reflection, and sense-making by visualizing learning activity traces.

Moreover, the output quality in RATsApp was found to be significantly influenced by the provision of formative feedback, which was highly valued by approximately 70\% of students. Around 60\% of students reported no issues with the quality of this feedback, indicating its usefulness and comprehensibility. Formative feedback, essential for enhancing student comprehension and skills~\citep{kulhavy1989feedback}, is typically delivered at the lecturer's discretion in traditional classrooms. However, in e-learning environments like RATsApp, the lack of face-to-face interaction poses a challenge for timely and effective feedback. Our system's expert-driven feedback design is based on the theoretical framework of \cite{hattie2007power} \citep[see also][]{wisniewski2020power}. This expert-driven approach was observed to be appealing to students in our study. Our findings suggest that this particular aspect of RATsApp was instrumental in positively influencing the perceived usefulness of the system, thereby increasing the students’ intention to use it.

Finally, the fact that a third of non-users express an intention to use the system, despite not currently doing so, implies that technical and access-related issues are likely the main obstacles to usage. According to TAM2, the high intention to use the system among non-users signals a perceived usefulness suggesting they recognize the app's potential value, despite not experiencing the system first hand. This is another indicator for external factors influencing the results. This emphasizes the necessity to simplify system access and reduce barriers such as VPN requirements. It also highlights the app's potential to draw more users once these external barriers are mitigated.

\subsection{Limitations}

\subsubsection{App Limitations}
The criteria list to estimate the skill level of the different competencies partly depends of the context of the RAT (e.g., completely new representations) as well as on some context independent difficulties of the RAT in regard to the competencies (e.g., information that is spread over several representations). Furthermore, the correctness of a solved RAT also depends on the general understanding of the targeted concept and not only on the competencies. These factors may distort the estimation of competencies and contribute to the fact that competence levels are only estimates.
The calculated level strongly depends on the given exercises and may not reflect a high absolute competence level. These limitations make it important to test how well our system estimates competencies and to have an exercise pool which fulfills diverse criteria of different difficulties.

\subsection{Future work}

In our study, we did not explore all potential factors related to the TAM2 framework that could contribute to the perceived usefulness of a system. Therefore, a natural extension of this research would be to examine the influence of these factors in the context of our AFS to identify any additional influential elements. Additionally, we aim to assess how these influential factors' impact can be extrapolated to various platforms. Furthermore, we are considering the implementation of additional surveys, such as the TAM3~\citep{venkatesh2008technology}, to further evaluate our system’s capabilities.

In light of our findings, we plan to take steps to increase the relevance of RATsApp to the student's study. Our initial strategy involves developing RATs that are more closely aligned with the lecture content and potentially to assessments. Furthermore, we are devising a strategy to underscore the advantages of the system to students, thereby highlighting its significance.

Despite our results suggesting that it may not be of significant importance,  we plan to enhance the user experience of RATsApp to encourage wider adoption and consistent usage. Our primary focus will be on improving navigation and accessibility. We will incorporate user feedback to refine the design and functionality of RATsApp, making it more intuitive and user-friendly. This could involve refining existing features or introducing new ones that better meet user needs. To address accessibility issues, we are contemplating the potential removal of the VPN requirement. 

Another significant area of focus is the refinement of our competence estimation. We plan to introduce weights for all criteria for each competence to reflect the fact that the importance of each criterion is different for the different competencies and another level of skill is needed. Furthermore, we want to weigh the influence of different RATs on the competence estimation. 
By giving more difficult RATs higher weights in the computation compared to easier RATs, as in RATs that can be answered correctly with a lower competence level, we aim to increase the precision of our competence estimation. In further studies, we will evaluate which criteria to keep, which to replace and which we need to add to effectively estimate the competence levels. Also, we want to investigate the influence of giving competence estimates as feedback on the final lecture grade. 

Moreover, we plan to integrate ML algorithms to predict students' scores in the final exams based on their performance. This could potentially help the lecturer as an alert system to provide additional support to students who need it. We also plan to use ML to suggest RATs to students according to the competence they need to strengthen.

In addition to the planned technical improvements, an ethical focus on positively shaping the practical field of learning and teaching will also continue. It is in our interest to go beyond a purely ethically unproblematic design of the system and data security concerns, and to broaden the view to the actors involved in the learning environment and their relation to each other. As mentioned earlier, we also want to consider the results that are relevant from an ethical perspective. This includes, among other things, the technical optimization of the RATsApp regarding the enhancement of interpersonal discourse. Further development should prioritize interaction between students and/or between students and instructors. One potential approach to facilitate this would be to establish learning groups for students and/or teaching-learning groups where students can engage in collaborative dialogue regarding their thoughts, challenges, and feedback \citep{solanki2019success, wilson2015belonging, stadtfeld2019integration}.

Furthermore, as it is possible that students may be unable to answer RATs beyond their existing learning competencies – whether due to perceived pressure to solve tasks correctly, anxiety related to exams, or lack of self-confidence – it is both ethically and didactically necessary to critically reflect on these aspects. Therefore, if required, it may be helpful to provide or facilitate interpersonal communication options, in addition to technical options and support, helping to overcome these challenges. This could potentially reduce the barriers between students and lecturers, for example, by posing a question anonymously, or by placing service offerings related to exam anxiety or other relevant topics in the RATsApp layout.

It also is imperative to address the potential for failure due to a mismatch between the student’s abilities and their chosen career path. In the event that a student consistently provides incorrect responses to tasks, a safety mechanism should be implemented. This could involve providing students with the option to directly consult with a lecturer and/or student counselor for interpersonal feedback and assistance in developing an accurate self-assessment. If this feedback is delivered in an honest and appreciative manner, a potential failure in the examination may present an opportunity for the student to explore alternative educational pathways that better align with their individual strengths and interests.

\section {Conclusion}\label{conclusion}

The creation of RATsApp - an open source AFS - is an interdisciplinary effort between physics and mathematics education researchers, engineers, computer scientists, and ethicists. For the development, we have considered several relevant research results on formative feedback, formative assessment, student and teacher dashboards, scaffolding support, STEM skill level estimation, and learning analytics. As STEM skills are at the forefront of today's challenges, there is a high demand for personalized learning tools from students who need more support outside of school. This was confirmed by our usability test, where around 70\% of students rated their intention to use as high.

Our study found that the survey results largely align with the predictions of the TAM2 framework. We observed a significant linear relation between the intention to use and the frequency of usage, emphasizing the importance of fostering usage intention. While ease of use correlates with perceived usefulness, it does not significantly correlate with the intention to use. However, perceived usefulness does significantly correlate with intention to use, output quality, and study relevance. This suggests that while certain areas, particularly visual and navigational elements, could benefit from refinement, interface enhancements may not necessarily lead to a substantial increase in usage. Instead, it might be more effective to align RATsApp with lecture content and exam material to enhance its relevance to the study. Furthermore, improving the output quality, especially for critical features like STEM competency and the dashboard, could have a positive impact on usage.

The survey results are promising overall. Most students responded positively to the expert-generated feedback, and a majority find the system beneficial. The positive feedback from students underscores the promising future of our application in supporting and facilitating learning. Therefore, addressing the issues mentioned above could further enhance the perceived usefulness and ease of use of RATsApp, thereby maintaining high system usage.

On a final note, it's crucial to understand that the usability survey primarily reflects students' perceptions and may not directly quantify the effectiveness of learning. For a more precise evaluation of learning progress, a comprehensive study could be conducted. This could involve pre- and post-tests to measure knowledge acquisition, monitoring individual student progress over time, or even incorporating a control group for comparison. Despite these considerations, our innovative approach of integrating formative feedback with STEM skill level estimation has demonstrated potential and is well-equipped to assist students in their academic journey.

We are committed to helping students and teachers in this fast-growing era of technology by creating an accessible tool for education that makes learning more effective – ethically speaking – more sustainable. Therefore, we have made RATsApp publicly available and invite contributions to its ongoing development. The source code can be accessed on GitLab: \url{https://gitlab.rhrk.uni-kl.de/ki4tuk/ratsapp}. This manuscript describes version 1.0 of the project. 

We wish to highlight the critical role that open-source projects like RATsApp play in enhancing education by providing accessible tools for learning that can be freely used, modified, and shared by anyone. Furthermore, RATsApp can also serve as a valuable research tool, enabling researchers to evaluate educational frameworks. Such projects promote collaboration and innovation while ensuring transparency and sustainability.

\backmatter

\bmhead{Acknowledgments}

We thank the lecturers who supported the implementation of the system: Dr. Florentine Kämmerer, Dr. Jean-Pierre Stockis, Prof. Dr. Jan von Delft, Prof. Dr. Vladislav Yakovlev, Dr. Ioachim Pupeza, and Dr. Jonathan Bortfeld. 

 \section*{Declarations}


 \begin{itemize}
 \item Funding: This work was supported by BMBF within the project KI4TUK.
 \item Availability of data and materials:The datasets generated during and/or analysed during the current study are available from the corresponding author on reasonable request.
 \end{itemize}



\begin{appendices}

\section{Criteria to estimate the STEM competencies}\label{secA}

As outlined in the main text, our objective is to assess three fundamental STEM competencies: mathematical literacy, data literacy, and representational competence. To achieve this, we have developed a list of objective criteria for RATs, as presented in Table~\ref{Table_SM}, each of which is associated with one or more competencies and can be utilized to determine the requisite skill level for each competence category in a RAT. Our list is derived from the criteria established by~\cite{scheid2017erhebung}.

\begin{table}
\renewcommand{\arraystretch}{1.1}
    \centering
\caption{This table presents a list of criteria derived from~\cite{scheid2017erhebung}. It also indicates the competencies to which each criterion is associated. In RATsApp, the list is implemented in German and has been translated into English for presentation here.\label{Table_SM}
}
\begin{tabular}{|p{0.45\linewidth}||>{\centering\arraybackslash}p{0.15\linewidth}|>{\centering\arraybackslash}p{0.2\linewidth}|>{\centering\arraybackslash}p{0.15\linewidth}|}
\hline
\textbf{Criteria}	& \textbf{Data Literacy} & \textbf{Representational Competency} & \textbf{Mathematical Literacy}\\ \hline
It must be assessed whether the given data are complete and free of errors (also with regard to possible erroneous conclusions and the applicability of analysis procedures and algorithms).
& x	& ~ & ~\\ \hline
An analysis procedure must be applied.
& x	& ~ & ~\\ \hline
A suitable visualization for the data must be chosen.
& x	& x & ~\\ \hline
A conclusion must be reached using (the visual representation of) data (e.g., based on correlations/connections in the data) or a mathematical result.
& x	& ~ & ~\\ \hline
A decision must be made based on analysis or statistics.
& x	& ~ & ~\\ \hline
The consequences of the mathematical result for the real problem must be recognized.
& ~	& ~ & x\\ \hline
A suitable mathematical description of the problem must be chosen. Here, mathematical concepts must be taken into account and suitable assumptions must be made.
& x	& x & x\\ \hline
Mathematical definitions have to be used.
& ~	& ~ & x\\ \hline
Mathematical formalism's have to be used.
& ~	& ~ & x\\ \hline
Mathematical algorithms have to be used.
& ~	& ~ & x\\ \hline
A calculation must be performed.
& ~	& ~ & x\\ \hline
The limitations of the mathematical model used must be recognized.
& x	& ~ & x\\ \hline
The relevant variables in a problem must be identified (necessary, among other things, to simplify a problem).
& x	& ~ & x\\ \hline
A suitable solution strategy must be selected from a list.
& ~ & ~ & x\\ \hline
In addition to the question, there is also task-relevant text.
& ~	& x & ~\\ \hline
In addition to the question, there is at least one picture.
& ~	& x & ~\\ \hline
In addition to the question, there is at least one diagram (e.g. graphs, tables, vector fields).
& ~ & x & ~\\ \hline
In addition to the question, there is at least one formula or mathematical symbol relevant to the task.
& ~	& x & ~\\ \hline
At least one connection is required between the forms of representation used.
& ~	& x & ~\\ \hline
The distribution of information over the involved forms of representation is partly redundant.
& ~ & x & ~\\ \hline
At least one used representation has a high degree of abstractness.
& ~	& x & ~\\
\hline
\end{tabular}
\end{table}

\section{Usability questionnaire}\label{secA1}

In the main text (see Fig. \ref{fig:questionnaire_category_bar_plot}), we presented aggregate results for each category from the Likert scale questionnaire (see Section~\ref{Usability}). Table~\ref{tab:used_questionnaire} shows these categories and their individual items. The numbers of each item in a category correspond to those shown in Fig. \ref{app_fig:questionnaire_frequency_bar_plot}. The complete questionnaire comprised three components: multiple-choice questions, a Likert-scale questionnaire, and written feedback, as detailed in Table~\ref{tab:used_questionnaire}. This table presents both the original questions, formulated in German, and their corresponding translations.

\begin{table}
\caption{Questions included in the Likert scale questionnaire presented in Section~\ref{Usability}.}\label{tab:used_questionnaire}
    \centering
    \begin{tabular}{p{0.25\linewidth} @{\hspace{.1cm}} p{0.75\linewidth} }
     \toprule\\[-0.1cm]
    Category & Items \vspace{1mm}\\ \midrule \\[-0.3cm]
     Intention to Use &1) I plan to use RATsApp, if I can access it.\\
      & 2) I think it is likely that I will use RATsApp when I can access it.\\\\ 
      Output Quality &1) The quality of the explanations of the different answer options is high. \\ 
    &2) I have no problem with the quality of the explanations to the different answer choices. \\ 
    &3) The quality of the estimation of my competence level is high.  \\ 
    &4) I have no problem with the estimation of my level of competence. \\
    &5) The quality of the display of the temporal progression of my answer correctness is high. \\ 
    &6) I have no problem with the quality of the display of the temporal progression of my answering skill.  \\\\
    Relevance to the Study &1) Using RATsApp is important for my studies.\\ 
    &2) The use of RATsApp is relevant to my studies.\\\\
    Perceived Ease of Use & 1) It is clear and understandable how I need to interact with RATsApp.\\ 
      &2) I don't have to think much about how to use RATsApp.\\ 
      &3) I find RATsApp is easy to use.\\ 
    &4) I find it easy to get RATsApp to do what I want.\\\\
      Perceived Usefulness &1) Using RATsApp will improve my performance in my studies.\\ 
      &2) By using RATsApp, I learn more.\\ 
      &3) By using RATsApp, I learn more effectively.\\ 
      &4) I find RATsApp useful for my studies.\\\\ 
      
    \botrule
    \end{tabular}
    
\end{table}

\begin{table}
\renewcommand{\arraystretch}{1.2}

\caption{Usability questionnaire used to evaluate RATsApp\label{tab:used_questionnaire}} 
    \centering
    \begin{tabular}{p{0.4\linewidth} @{\hspace{.3cm}} p{0.42\linewidth} @{\hspace{.3cm}} p{0.2\linewidth} }
     \toprule\\[-0.25cm]
English translation & German original & Input type  \\ \midrule \\[-0.3cm]
I attended the following lecture & Ich habe folgende Vorlesung besucht  & Multi select field\\
I attended the following 2nd lecture\textsuperscript{1} & Ich habe folgende 2. Vorlesung besucht  & Multi select field\\
I attended the following 3rd lecture\textsuperscript{1} & Ich habe folgende 3. Vorlesung besucht  & Multi select field\\
I attended the following 4th lecture\textsuperscript{1} & Ich habe folgende 4. Vorlesung besucht  & Multi select field\\
I attended the following 5th lecture\textsuperscript{1} & Ich habe folgende 5. Vorlesung besucht  & Multi select field\\
How old are you? & Wie alt sind Sie? & Number field\\
What is your gender? & Was ist Ihr Geschlecht? & Multi select field\\
Do you have a RatsApp account? &  Haben Sie einen RatsApp Account? & Multi select field\\
How often do you use RatsApp? & Wie oft nutzen Sie RatsApp? & Multi select field\\
Do you have any general comments about RatsApp? &  Haben Sie allgemeine Anmerkungen zu RatsApp? & Free text field\\
I plan to use RATsApp, if I can access it. & Ich habe vor RatsApp zu nutzen, wenn ich darauf zugreifen kann. & 7 Point Likert Scale\\
I think it is likely that I will use RATsApp when I can access it. & Ich halte es für wahrscheinlich, dass ich RatsApp nutzen werde, wenn ich darauf zugreifen kann. & 7 Point Likert Scale\\
Do you have any other comments on your intention of use? & Haben Sie weitere Anmerkungen zur geplanten Nutzung? & Free text field\\
Using RATsApp will improve my performance in my studies. & Die Nutzung von RatsApp verbessert meine Leistungen im Studium. & 7 Point Likert Scale\\
By using RATsApp, I learn more. & Durch die Nutzung von RatsApp lerne ich mehr. & 7 Point Likert Scale\\
By using RATsApp, I learn more effectively. & Durch die Nutzung von RatsApp lerne ich effektiver. & 7 Point Likert Scale\\
I find RATsApp useful for my studies. & Ich finde RatsApp für mein Studium nützlich. & 7 Point Likert Scale\\
Do you have any other comments about the usefulness of RatsApp? & Haben Sie weitere Anmerkungen zur Nützlichkeit von RatsApp? & Free text field\\
It is clear and understandable how I need to interact with RATsApp. & Es ist klar und verständlich wie ich mit RatsApp interagieren muss. & 7 Point Likert Scale\\
I don't have to think much about how to use RATsApp. & Ich muss nicht viel über die Benutzung von RatsApp nachdenken. & 7 Point Likert Scale\\
I find RATsApp is easy to use. & Ich finde RatsApp ist einfach zu benutzen. & 7 Point Likert Scale\\
I find it easy to get RATsApp to do what I want. & Ich finde es einfach RatsApp dazu zu bringen das zu tun was ich möchte. & 7 Point Likert Scale\\
Do you have any other comments about the usability of RatsApp? & 
Haben Sie weitere Anmerkungen zur Bedienbarkeit von RatsApp? & Free text field\\
Using RATsApp is important for my studies. & Die Nutzung von RatsApp ist wichtig für mein Studium. & 7 Point Likert Scale\\
The use of RATsApp is relevant to my studies. & Die Nutzung von RatsApp ist relevant für mein Studium. & 7 Point Likert Scale\\
Do you have any other comments on the relevance of RatsApp? & Haben Sie weitere Anmerkungen zur Relevanz von RatsApp? & Free text field\\
The quality of the explanations of the different answer options is high. & Die Qualität der Erklärungen zu den verschiedenen Antwortmöglichkeiten ist hoch. & 7 Point Likert Scale\\
    \botrule
\end{tabular}
\noindent{\footnotesize{\textsuperscript{1} The question was displayed only if the previous question was answered.}}
\end{table}




\end{appendices}
\clearpage


\bibliography{citations}


\end{document}